\documentclass[aps,amsmath,showpacs,showkeys,superscriptaddress,twocolumn]{revtex4}
\usepackage{color}
\usepackage{graphicx}
\usepackage{amssymb}
\usepackage{subfigure}
\usepackage{titlesec}
\begin{document}
\title{Travelling wave solutions of equations in the Burgers Hierarchy}
\author{ Amitava Choudhuri}
\affiliation{Department of Physics, The University of Burdwan, Purba Bardhamam, 713104,  West Bengal, India}
\author{Madan Mohan Panja}
\affiliation{Department of Mathematics, Visva-Bharati University, Santiniketan, 731235, West Benga, India}
\author{Supriya Chatterjee}
\affiliation{Department of Physics, A. P. C. Roy Government College, Himanchal Vihar, Matigara, Siliguri, 734010, West Bengal, India}
\author{Benoy Talukdar}
\affiliation{Department of Physics, Visva-Bharati University, Santiniketan, 731235, West Benga, India}
\email{binoyt123@rediffmail.com}
\begin{abstract}
We emphasize that construction of travelling wave solutions for partial differential equations is a problem of considerable interest and thus introduce a simple algebraic method to generate such solutions for equations in the Burgers hierarchy. Our method based on a judicious use of the well known Cole-Hopf transformation is found to work satisfactorily for higher Burgers equations for which the direct method of integration is inapplicable. For Burgers equation we clearly demonstrate how does the diffusion term in the equation counteract the nonlinearity to result in a smooth wave. We envisage a similar study for higher equations in the Buggers hierarchy and establish that (i) as opposed to the solution of the Burgers equation, the purely nonlinear terms of these equations support smooth solutions and more interestingly (ii) the complete solutions of all higher-order equations are identical.
\end{abstract}
\maketitle
\section*{1. Introduction}
In physical theories both linear and nonlinear partial differential equations are used to study wave propagation. A wave is a recognizable signal that transfers energy from one part of a medium to another part with a speed of propagation. The simplest form of a wave is given by a mathematical function of the form
\begin{equation}
    u(x,t)=f(x-ct),
\end{equation}
where $x$ stands for the space coordinate and $t$, the time. The wave as written in Eq.(1) is qualified as a travelling wave with $\xi=x-ct$, the so-called  travelling coordinate. It propagates with speed $c>0$ along the x-direction and satisfies the constant coefficient transport equation, $u_t+cu_x=0$. At $t=0$ the wave has the form $f(x)$ representing the initial profile. Thus $f(x-ct)$ stands for the profile at time $t$ and gives the initial profile translated to the right by  $ct$ spatial units.
\par One fundamental question in the theory of partial differential equations (PDEs) is that whether a given PDE with prescribed initial and boundary conditions admits a travelling wave solution .This point has been beautifully expounded in the work by Bejeia et al. \cite{1}. Moreover, construction of travelling wave solutions for physically important families of partial differential equations has always been regarded as a problem of interesting curiosity. For example, Sinuvasan  et al. \cite{2} made use of an approach based on Lie point symmetries to obtain travelling wave and self-similar solutions of equations in the Burgers hierarchy \cite{3}. Self-similar solutions \cite{4} refer to mathematical solutions to equations when the solutions remain unchanged under scaling of space and time variables. For example if $u(x,t)$ is a solution of the diffusion (heat) equation $u_t=ku_{xx}$, then so is the rescaled function $\nu(x,t)=u(\sqrt{ax},at)$. This property makes them extremely useful to simplify complex problems in physics and other related fields. The object of the present work is, however, to provide an uncomplicated method to construct only travelling-wave solutions for equations in the Burgers hierarchy. In the course of our study we shall see that the approach developed by us, on the one hand, provides a useful supplement for more advanced studies \cite{2,5} and, on the other hand, can be used for class-room demonstration of the topic to graduate and under-graduate students of  physics and mathematics. In the next section we introduce Burgers equation and equations in the Burgers hierarchy, and make some useful comments. In Sect.3 we first make use of a straightforward method of integration to obtain the travelling wave solution of Burgers equation and then introduce a linearization technique of the equation which provides a useful basis to integrate it. For pedagogic purpose we study the interplay between dissipative and nonlinear effects, which converts shock discontinuities to well behaved solutions. Problems for solving  higher-order equations in the Burgers hierarchy are considered in Sect.4. Here we observe that  the method of direct integration is not applicable for these equations while our method based on  linearization of the equations works satisfactorily for all members in the hierarchy. In close analogy with the treatment of Burgers equation here we also study the interplay between effects of linear and nonlinear terms on propagation of travelling  waves and make some crucial observation. Finally, in Sec.5 we summarize our outlook on the present work and make a few concluding remarks. 
\section*{2. Equations in Burgers Hierarchy}
For a wave represented by $u=u(x,t)$ Burgers equation \cite{6} reads
\begin{equation}
    u_t=2uu_x+\nu u_{xx},
\end{equation}
with $\nu$ the kinematic viscosity of the medium in which the wave propagates. Equation (2) provides a fundamental partial differential equation that accounts for the combined effects of nonlinearity and diffusion in fluid flow and plays a special role to study the interplay between convection and diffusion, which are two crucial aspects of fluid dynamics. In writing Eq.(2) we used subscripts to denote partial derivatives with respect to time and space variables, in particular, $u_{xx}=\frac{\partial^2u}{\partial x^2}$. Clearly, the nonlinearity of the equation arises through the term $uu_x$while the term $u_{xx}$ accounts for viscous effect of the fluid.
\par The equations in the Burgers hierarchy are given by \cite{3}
\begin{equation}
    u_t=\partial_x((\partial_x+u)^nu)=0,\;\;\;n=1,2,3,....
\end{equation}
where $\partial_x=\partial/\partial x$. Clearly for $n=1$ in Eq.(3) gives Burgers equation provided we let $u_{2x}\rightarrow \nu u_{2x}$. The presence of $\nu$ in the Burgers equation allows us to control the nonlinear wave motion by the effect of viscosity in the medium. The second, third and fourth members of the hierarchy as obtained from Eq.(3) are given below.    
\begin{equation}
    u_t=3u^2u_x+3u_x^2+3uu_{2x}+\nu u_{3x},
\end{equation}
\begin{equation}
    u_t=4u^3u_x+12uu_x^3+6u^2u_{2x}+10u_xu_{2x}+4uu_{3x}+\nu u_{4x},
\end{equation}
and
\begin{equation}
\begin{split}
u_t=5u^4u_x+30u^2u_x^2+15u_x^3+10u_{2x}^2+10u^3u_{2x}\\+50uu_{x}u_{2x}+10u_{2x}^2+10u^2u_{3x}+15u_xu_{3x}\\+5uu_{4x}+\nu u_{5x}.
\end{split}
\end{equation}
In writing Eqs.(4)-(6) we allowed $u_{nx}=\nu u_{nx}$ for $n=2,3$ and 4 respectively. Looking at these equations we see that the order of the dissipative-like term/order of the differential equation increases as we go along the hierarchy. At the same time we have rapid increase in the number of nonlinear terms. As with Eq.(1)  higher-order Burgers equations are also used to model various physical phenomena including wave propagation, shock waves and turbulence.
\section*{3. Travelling wave solution of Burgers equation}
To find the travelling-wave solution of the Burgers equation we substitute $u(x,t)$ from Eq.(1) in Eq.(2) to get
\begin{equation}
    cf'(\xi)+2f(\xi)f'(\xi)+\nu f''(\xi)=0.
\end{equation}
On integration Eq.(7) gives
\begin{equation}
    d+cf(\xi)+f^2(\xi)+\nu f'(\xi)=0
\end{equation}
where $d$ is a constant of the integration. Interestingly, the above first-order nonlinear differential equation can be solved to obtain travelling wave solution of the Burgers equation in the form
\begin{equation}
f(\xi)=\frac{1}{2}(-c+\sqrt{c^2-4d}\tanh(\frac{1}{2}\sqrt{c^2-4d})(c_1\nu-\xi))
\end{equation}
with $c_1$, a new constant of integration. 
\par In the following we now introduce an alternative method to obtain the solution in Eq.(8). To that end we make use of a Cole-Hopf like transformation \cite{7,8}
\begin{equation}
f(\xi)=\frac{g'(\xi)}{g(\xi)}
\end{equation}
to reduce Eq.(8) in the form
\begin{equation}
g(\xi)(\gamma g''(\xi)+cg'(\xi))-(\nu-1)g'^2(\xi)+dg^2(\xi)=0
\end{equation}
which, unfortunately, is a nonlinear differential equation. Despite that it can be solved analytically to get
\begin{equation}
g(\xi)=c_2e^{\frac{1}{2}}q(\xi)
\end{equation}
with
\begin{equation}
q(\xi)=c\xi+2\nu\ln(\cos(\frac{\sqrt{4d-c^2}}{2\nu})-\frac{1}{2}c_1\sqrt{4d-c^2}).
\end{equation}
Interestingly, Eq.(12) and its first derivative when used in Eq.(10) gives the same result for the solution $f(\xi)$ as give in Eq.(9). Understandably, $c_2$ in Eq.(12) is a new constant of integration.
\par In many advanced studies \cite{9,10} Burgers equation are often studied  for $\nu=1$. In this special case the above method  to compute $f(\xi)$ using  Eq.(10) takes a very simple form. For example, in the case of Burgers equation $g(\xi)$ satisfies the simple differential equation
\begin{equation}
g''(\xi)+cg'(\xi)+dg(\xi)=0.
\end{equation}
It is straightforward to use the solution of Eq.(14) to obtain the result in Eq.(9) for $\nu=1$.
\par In the next section we shall see that, as opposed to Eq.(8), higher-order Burgers equations lead to  equations which are of order $\geq2$. These equations cannot be solved analytically such that the method of direct integration fails for all higher-order equations in the hierarchy.  But our method based on Cole-Hopf transformation works satisfactorily to solve all such equations. Meanwhile, let us study the interplay between nonlinearity and viscosity on the solution of the Burgers equation.
\par Without the viscous term  the solution of Eq.(2) allows shocks to be formed due to squeezing effect of the nonlinear terms and finally breaks down. The presence of the viscous /diffusion term prevents gradual distortion of the wave by counteracting the effect of nonlinearity. In order to demonstrate this we display in Fig.1 the travelling wave $f(\xi)$ from Eq.(9) as a function of the coordinate $\xi$ for three different values of the kinetic viscosity, namely, $\nu=$0.01, 0.5 and 1.0 respectively.  Here we have chosen to work with the constant of integration $c_1=1$, $c=1$ and $d=0$. The wave profile $f(\xi)$ moves  from left to the right unchanged in form. The solid line in the figure  gives the curve for $\nu=0.01$. In this case the effect of the viscous term  on the solution is almost negligible such that the travelling wave solution, as represented by the solid line, is a step function characteristic of the shock wave solution of the inviscid Burgers equation. The dashed line with giving variation of $f(\xi)$ as a function of $\xi$ for $\nu=0.5$   clearly shows how does the effect of viscosity smooth out the shock discontinuity. The dot-dashed curve represents the travelling wave for $\nu=0.5$ having maximum deviation from the curve for inviscid solution. 
\begin{figure}
    \centering
    \includegraphics[width=0.5\linewidth]{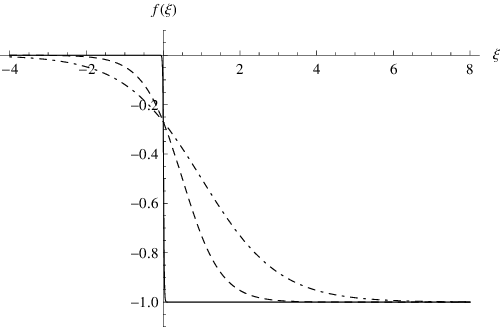}
    \caption{Solutions of Burgers equation as a function of the travelling coordinate  for three different values of the kinetic viscosity.}
\end{figure}
\section*{4. Travelling wave solutions of higher Burgers equations}
In close analogy with the treatment presented for the Burgers equation we first try to construct an analytic result for the travelling wave of the first member in the Burgers hierarchy as given in Eq.(4). In the travelling coordinate the equation of our interest reads
\begin{equation}
\nu f'''(\xi)+3f(\xi)f''(\xi)+3f'^2(\xi)+3f^2(\xi)f'(\xi)+cf'(\xi)=0.
\end{equation}
Eq.(15) can be integrated to get
\begin{equation}
\nu f''(\xi)+3f(\xi)f'(\xi)+f^3(\xi)+cf(\xi)+d=0
\end{equation}
with $d$, a constant of the integration. The first integral of Burgers equation as given in Eq.(8) could be solved analytically to find the associated travelling wave solution. But unfortunately, Eq.(16) cannot be solved similarly to construct the travelling wave solution of Eq.(15). To provide an analytic solution for the problem we thus take recourse to the use of our ansatz  in Eq.(10) which when used in Eq.(16) gives 
\begin{equation}
\begin{split}
g^2(\xi)(\nu g'''(\xi)+cg'(\xi))-3(\nu-1)g(\xi)g'(\xi)g''(\xi)\\+2(\nu-1)g'^3(\xi)+dg^3(\xi)=0.
\end{split}
\end{equation}
Eq.(17) represents a very complicated third order nonlinear differential equation which cannot be solved analytically. But interestingly enough for $\nu=1$ it takes a very neat form given by
\begin{equation}
g'''(\xi)+cg'(\xi)+dg(\xi)=0.
\end{equation}
For simplicity of presentation henceforth we shall take $d=0$. The result for solution of Eq.(18) for $d\neq 0$ is given in Appendix A. Making use of the solution of Eq.(18) the travelling wave solution of Eq.(4) for $\nu=1$ and $c=1$ can be written in a very simple form 
\begin{equation}
f(\xi)=\frac{\cos\xi+\sin\xi}{1-\cos\xi+\sin\xi}.
\end{equation}
We now proceed to obtain a solution of Eq.(4) in the absence of the term ($\nu=0$). Understandably, this solution is due to nonlinear terms only.   In this case  Eq.(16) gives
\begin{equation}
3f'(\xi)+f^2(\xi)+c=0
\end{equation}
which, in view of Eq.(10), reads 
\begin{equation}
3g(\xi)g''(\xi)-2g'^2(\xi)+cg^2(\xi)=0.
\end{equation}
The solution of Eq.(21) can be written as
\begin{equation}
g(\xi)=\cos^3(\frac{\sqrt{c}}{3}(\xi-3))
\end{equation}
which via Eq.(10) leads to the travelling wave solution of Eq.(4) in absence of the term $u_{3x}$ in the form
\begin{equation}
f(\xi)=-\tan(\frac{1}{3}(\xi-3))
\end{equation}
for $c=1$.
We noted that, in the absence of the viscous term ($u_{2x}$), Burgers equation allows shocks to be formed due to nonlinearity in the medium and demonstrated how breaking of the shock wave is counteracted by the presence of viscosity. The expressions in Eq.(19) and Eq.(23) represent the travelling waves of the first member of the Burgers hierarchy in the presence and absence of the linear term $u_{3x}$. As in the case of Burgers equation, it will be interesting to compare the spatial evolution of these two solutions with a view to gain some physical weight for the effect of the term $u_{3x}$ on the solution supported by the nonlinear terms only. Fig.(2) gives the plots of $f(\xi)$ as a function of $\xi$ for the expressions in Eqs.(19) and (23) respectively. The solid lines give the variation of $f(\xi)$ in Eq.(23) while the dashed lines portray their modification (Eq.(19)) by the influence of the term $u_{3x}$.    
\begin{figure}
    \centering
    \includegraphics[width=0.5\linewidth]{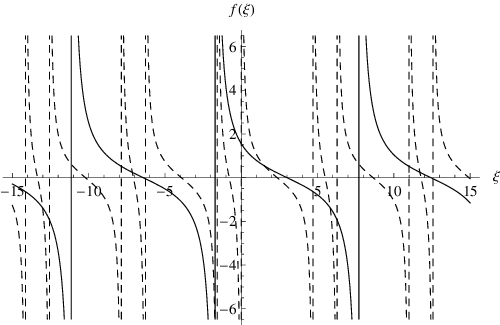}
    \caption{Solutions of the first equation in Burgers hierarchy as a function of the travelling coordinate. The solid and dashed lines represent results for equations in absence and presence of the term $u_{3x}$.}
\end{figure}
Looking closely into this figure we see that, as opposed to the shock wave in Fig.1, here we have smooth curves as solutions of Eq.(4) without the term $u_{3x}$.The dashed curves indicate that in the presence of this term and the solid line is replaced a number of lines. To clearly visualize how a typical solution of the equation without the term $u_{3x}$ is modified by the effect of the linear term we display in Fig.3 $f(\xi)$ in a very small interval of $\xi$. 
\begin{figure}
    \centering
    \includegraphics[width=0.5\linewidth]{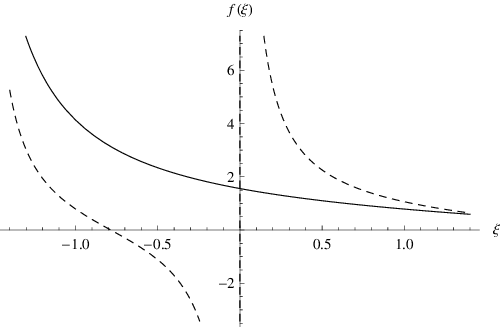}
    \caption{Solution of the first member of the Burgers hierarchy for $\xi$ in the interval $(-2,2)$. The solid and dashed curves carry the same meaning as those in Fig.2.}
\end{figure}
From this figure we see that a solution of the equation in the absence of the term $u_{3x}$ (solid line) modifies to three different solutions (dashed lines) in the presence of dissipative-like term. Amongst these solutions one is a vertical line while the other two lie symmetrically about the solid line.
\par In the above context it remains an interesting curiosity to see how the curves in Figs.2 and 3 change or modify as we go along the hierarchy. Interestingly, we have checked and verified that the solutions of higher-order equations do not exhibit any change in shape of the solutions than those exhibited by the solutions of Eq.(4). As a typical example in respect of this we provide below the solutions of Eq.(6) representing the fourth member in the hierarchy. The solution in the presence of the term $u_{5x}$ is given by
\begin{equation}
    f(\xi)=\frac{\sqrt{2}(e^{\sqrt{2}\xi}\cos\frac{\xi}{\sqrt{2}}-\sin\frac{\xi}{\sqrt{2}})}{-ae^{\frac{\xi}{\sqrt{2}}}+(1+e^{\sqrt{2}\xi})\cos\frac{\xi}{\sqrt{2}}+(1+e^{\sqrt{2}\xi})\sin\frac{\xi}{\sqrt{2}}}
\end{equation}
while the solution in the absence of the term $u_{5x}$ reads
\begin{equation}
    f(\xi)=\frac{\sqrt{2}(-1+e^{\sqrt{2}\xi})\cos\frac{\xi}{\sqrt{2}}-(1+e^{\sqrt{2}\xi})\sin\frac{\xi}{\sqrt{2}}}{ae^{\frac{\xi}{\sqrt{2}}}+2(1+e^{\sqrt{2}\xi})\cos\frac{\xi}{\sqrt{2}}}.
\end{equation}
For Eqs. (24) and (25) a plot similar to that in Fig.2 is presented in Fig.4.
\begin{figure}
    \centering
    \includegraphics[width=0.5\linewidth]{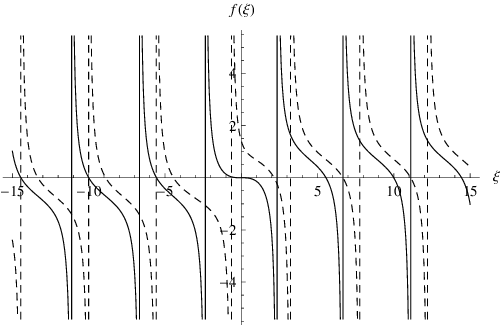}
    \caption{Solutions of the fourth equation in Burgers hierarchy as a function of the travelling coordinate . The solid and dashed lines represent results for equations in absence and presence of the term $u_{5x}$.}
\end{figure}
The curves in Fig.4 representing solutions of the fifth-order equation are similar to those in Fig.2 which displays  solutions of the third-order equation. It appears that the solutions in Fig.4 are closely spaced on the axis as compared to those in Fig.2. For example in the interval $-15\leq\xi\leq 15$ the number of  solid curves in Fig.2 is 4 while that in Fig.4 is 7. One can verify that the observed nature of the solutions do not change as one goes along the hierarchy of Burgers equation implying that the pattern of solutions of all higher-order equations is identical.
\section*{5. Concluding Remarks }
Burgers equation represents the simplest model for analyzing combined effect of nonlinear convection and diffusion. The presence of the viscous term helps suppress the wave-breaking and thus smooth out shock discontinuities to have a well-behaved smooth solution. The Burgers hierarchy, as introduced in Eq.(3) is a well known family of nonlinear evolution equations. Each partial differential equation in the hierarchy is characterized by a number of nonlinear terms and one linear space derivative term giving the order of the equation. 
\par We have begun this work by pointing out the importance of constructing travelling wave solutions of nonlinear partial differential equations and subsequently derived an uncomplicated method to construct such solutions for equations of the Burgers hierarchy. Our method proceeds by making use of Cole-Hopf transformation. We have clearly demonstrated how does the viscous effect modulated by the linear term influence the solution of Burgers equation supported by only the nonlinear terms. This point has extensively been studied in the mathematical literature. But it appears that similar studies have not yet been envisaged for higher-order equations in the Burgers hierarchy. We found that, as opposed to the solution of Burgers equation, the purely nonlinear terms in higher-order equations support continuous solutions each of which breaks into three components in the presence of linear terms ($u_{ns}$, $n=$3, 4 etc.). This is an invariant property of all equations in the Burgers Hierarchy. In view of this we feel that the work presented in this work provides a useful addendum to approaches followed in references 1 and 2 to find travelling wave solutions of nonlinear partial differential equations. Recently, it has been shown that soliton solutions of the KdV, mKdV and NLS equations can be constructed by going over to the travelling coordinates \cite{11}. Thus it will be interesting to look for travelling wave solutions of other families of physically interesting nonlinear differential equations.
\vskip0.25cm
{\bf Appendix A: The result for solution of Eq. 18 for the integration constant $d\neq 0$} 
\par The solution of Eq.(18) in which $d\neq 0$ can be written as
\begin{equation}
    g(\xi)=c_1e^{a\xi}+e^{b\xi}(c_2\cos\lambda\xi+c_3\sin\lambda\xi).\tag{A1}
\end{equation}
The values of $a$, $b$ and $\lambda$ in Eq.(A1) are as follows.
\begin{equation}
    a=(2^{\frac{1}{3}}\alpha^2-24^{\frac{1}{3}}c)/(6^{\frac{2}{3}}\alpha),
    \tag{A2}
\end{equation}
\begin{equation}
    b=(2\sqrt{3}c-2^{\frac{1}{3}}3^{\frac{1}{6}}\alpha^2)/(2^{\frac{5}{3}}2^{\frac{5}{6}}\alpha)
    \tag{A3}
\end{equation}
and 
\begin{equation}
    \lambda=(6c+2^{\frac{1}{3}}3^{\frac{2}{3}}\alpha^2)/(2^{\frac{5}{2}}3^{\frac{5}{6}}\alpha),
    \tag{A4}
\end{equation}
where 
\begin{equation}
    \alpha=(\sqrt{12c^2+81d^2}-9d)^{\frac{1}{3}}.
    \tag{A5}
\end{equation}
For $d=0$ and $c=1$, we get from Eq.(A5) $\alpha=(2\sqrt{3})^{\frac{1}{3}}$ such that $a=b=0$ and $\lambda=1$. Eq.(A1) then gives
\begin{equation}
    g(\xi)=c_1+c_2\cos\xi+c_3\sin\xi.
    \tag{A6}
\end{equation}
The above expression for $g(\xi)$ via Eq.(10)   gives the travelling wave solution in Eq.(19) provided we choose the constant of integrations as $c_1=c_3$ and $c_2=-1$. 


\vskip 0.5 cm
\begin{thebibliography}{10}
\bibitem{1} Bezeia D, Das Ashok, Losano L and Santos M J 2010 'Travelling wave solutions  of nonlinear partial differential equations', Applied Mathematics Letters {\bf 23} 681.
\bibitem{2} Sinuvasan R, Tamizhmani K M and Leach P G L 2017 'Symmetries, travelling wave and self-similar solutions of the Burgers hierarchy', Applied Mathematics and Computation {\bf 303} 165.
\bibitem{3} Kudryashov N A and Sinelshchikov D I  2009 'Exact solution of equations for the Burgers hierarchy', Applied Mathematics and Computation {\bf 215} 1293.
\bibitem{4} Gratton J 1991 'Similarity and self-similarity in fluid dynamics', Fundamentals of Cosmic Physics, Vol. 15 Gordon and Breach New York
\bibitem{5} Kudryashov N A 2021 'Generalized Hermite polynomials for the Burgers hierarchy and point vortices', Chaos, Solitins and Fractals {\bf 151} 111256
\bibitem{6} Burgers J M 1948 'A Mathematical model illustrating the theory of turbulence' Advances in Applied Mechanics {\bf 1} 171         
\bibitem{7} Cole J D 1951 'On a quasi-linear parabolic equation occurring in aerodynamics' Quarterly of Applied Mathematics {\bf 9} 225.
\bibitem{8} Hopf E 1950 'The partial differential equation $u_t + uu_x = u_{xx}$' Communications on Pure Appllied Mathematics { \bf 3}, 201.
\bibitem{9} Talukdar B, Ghosh S and Das U 2005 'Inverse variational problem and canonical structure of Burgers equations' J. Math. Phys. {\bf 46} 043506.
\bibitem{10} Cao X and Xu C 2010 'A Backlund Transformation for the Burgers hierarchy', Abstract and Applied Analysis {\bf 
 2010} 241898.
\bibitem{11} Supriya Chatterjee, Pranab Sarkar and Benoy Talukdar 2025 'Travelling wave solutions and Solitons of KdV, mKdv and NLS equations' arXiv:2508.18994v1 [math-ph] 26 Aug.
\end{thebibliography}
\end{document}